\documentclass{llncs}
\usepackage{graphicx, subfigure}
\usepackage{url}

\title{Cerulean: A hybrid assembly using high throughput short and long reads}

\author{Viraj Deshpande\inst{1} \and Eric DK Fung\inst{2}
Son Pham\inst{1} \and Vineet Bafna\inst{1}}

\institute{Department of Computer Science \& Engineering, University of California,\\ San Diego, CA, USA,
\email{vbafna@eng.ucsd.edu},
\and
Bioinformatics Undergraduate Program, Department of Bioengineering,\\University of California, San Diego, CA, USA}

\begin{document}

\maketitle
\begin{abstract}

Genome assembly using high throughput data with short reads, arguably, remains
an unresolvable task in repetitive genomes, since when the length of a repeat
exceeds the read length, it becomes difficult to unambiguously connect the
flanking regions.  The emergence of  third generation sequencing (Pacific
Biosciences) with long reads enables the opportunity to resolve complicated
repeats that could not be resolved by the short read data. However, these long
reads have high error rate and it is an uphill task to assemble the genome
without using additional high quality short reads. Recently, Koren et al.
2012~\cite{koren2012hybrid} proposed an approach to use high quality short
reads data to correct these long reads and, thus, make the assembly from long
reads possible. However, due to the large size of both dataset (short and long
reads), error-correction of these long reads requires excessively high
computational resources, even on small bacterial genomes. In this work, instead
of error correction of long reads, we first assemble the short reads and later
map these long reads on the assembly graph to resolve repeats.

\textbf{Contribution}:  We present a hybrid assembly approach that is
both computationally effective and produces high quality assemblies. Our
algorithm first operates with a simplified version of the assembly graph
consisting only of long contigs and gradually improves the assembly by adding
smaller contigs in each iteration. In contrast to the state-of-the-art long
reads error correction technique, which requires high computational resources
and long running time on a supercomputer even for bacterial genome datasets, our
software can produce comparable assembly using only a standard desktop in a
short running time.

\end{abstract}

\section{Introduction}

\begin{figure}[tb]
 \includegraphics[width=\textwidth]{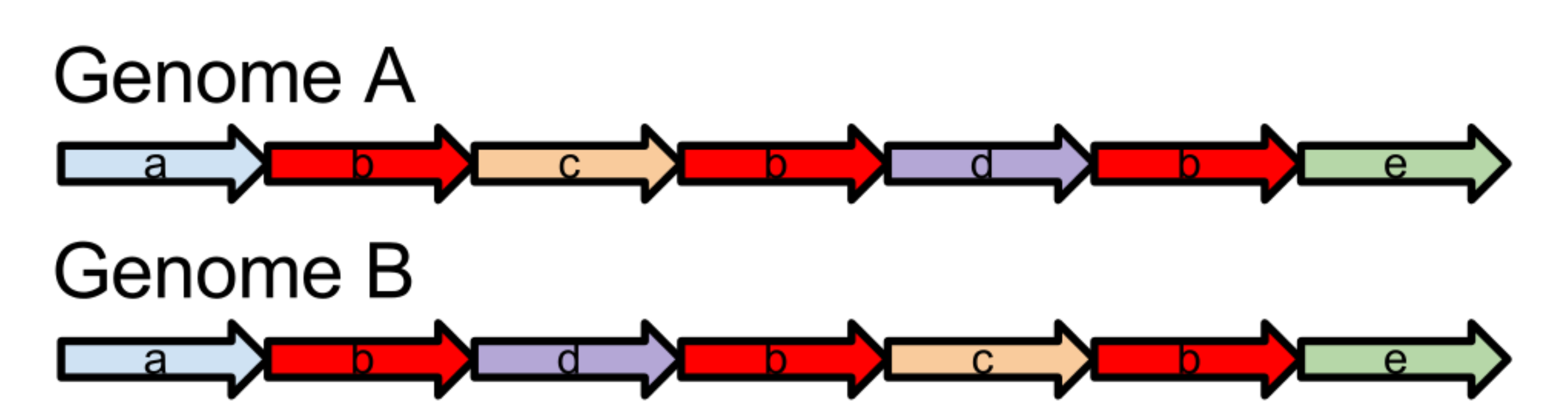}
 \caption{Two possible genomic architectures, Genome A and Genome B each with a
large repeat b. Both putative genomes can generate the same set of paired-end
reads. With the information from paired-end reads, the assembler can only
identify shorter contigs a, b, c, d and e but not the entire true genome.}
 \label{fig:unspannedrepeats}
\end{figure}

The advent of high throughput sequencing technologies has generated a lot of
interest from the computational perspective of de novo assembly of genomic
sequences. A major breakthrough in the massively parallel high-throughput
sequencing technologies includes the second generation sequencing platforms
including those from Illumina and Life Technologies. These platforms generate
paired-end reads with length of the order of 100 or 250 base pairs. The mean
end-to-end distance between the paired-end reads, called the \textit{insert
size}, is around 300 to 500 base-pairs. The paired-end reads can be sequenced
with high accuracy and high depth of coverage. Two approaches are broadly used
by assembly tools to assemble the paired-end reads into complete genomes of
genomic contigs: (i) Overlap-layout-consensus
assembly was introduced by Staden~\cite{staden1979strategy}, which was later
improvised by the introduction of string graph
by Myers~\cite{myers2005fragment} (Celera Assembler~\cite{myers2000whole},
SGA~\cite{simpson2012efficient}); (ii) de Bruijn graph-based assembly was
originally proposed by Idury and Waterman~\cite{idury1995new} and extended by
Pevzner et al~\cite{pevzner2001eulerian} (Euler-SR~\cite{chaisson2008short},
ABySS~\cite{simpson2009abyss}, Velvet~\cite{zerbino2008velvet}). These de
novo assembly approaches can generate high quality assemblies using the high
quality paired-end reads. However, these paired-end reads are unable to span
large repeats and as a result the assembled contigs are often short. Figure
\ref{fig:unspannedrepeats} shows an example where the assembler is
unable to identify the true genome architecture using only short read.

Newer high throughput sequencing platforms~\cite{eisenstein2013companies}
target the length limitation of second generation short reads by generating
libraries with a large span. The prominent technologies include: (i) jumping
libraries which generate small mate-pair reads of around 150 base pairs and
variable span of 5 kbp or more; (ii) long reads from Pacific Biosciences with
variable length ranging from 1 kbp to 20 kbp and (iii) genomic fragments
amplified by Moleculo technology (size: 1.5 kbp to 15
kbp~\cite{waldbieser2013production}) and then sequenced using short read
sequencing technologies. In this work we focus on PacBio long reads generated by
directly sequencing entire genomic fragments. In the rest of this article we
will use the term \textit{short reads} to describe the paired-end reads from
Illumina and \textit{long reads} to describe reads generated using the
Pacific Biosciences RS platform.

Long reads generated from PacBio RS can easily span most repeats and have the
potential to produce very large assembled contigs. Unfortunately, these  reads
can have a very high error rate with mean error rate as high as 16\%. Hence they
are difficult to assemble by themselves and require very high coverage; the
assembly quality falls rapidly with smaller
coverage~\cite{chin2013nonhybrid}. However, combining the high quality of
second generation short reads and the large length of the long reads, these
datasets can be processed simultaneously to produce very long genomic contigs
that otherwise required costly low-throughput techniques. 

Recent efforts (PacbioToCA~\cite{koren2012hybrid},
LSC~\cite{au2012improving}) have focused on mapping short reads to the
erroneous long reads to correct the long reads using aligners like
NovoAlign~\cite{hercus2009novocraft} and GMAP~\cite{wu2005gmap} which can
allow large edit distance for the mapping. These corrected long reads are then
used to generate an assembly with significantly longer contigs. However, such
mapping from all short reads to all long reads with large edit-distance is
computationally expensive and requires a large running time even for small
bacterial datasets. Furthermore, if there are two or more similar regions in the
genome, the short reads from one region can still map to long reads from the
other region given the high edit-distance. Reads corrected in such fashion may
create spurious adjacencies leading to misassemblies.

An alternative approach is to first assemble the high coverage short read
dataset to produce high quality contigs and use long reads for scaffolding.
Previous tools that use this approach include AHA
scaffolder~\cite{bashir2012hybrid} and
ALLPATHS-LG~\cite{ribeiro2012finished}. However, these approaches are
specialized to perform hybrid assembly in the presence additional libraries
including Roche 454 reads for AHA scaffolder and jumping libraries of mate-pair
reads for ALLPATHS-LG. 
Following this approach, a hybrid assembler will essentially take as input:  the
assembler should find the correct traversal of genome on the graph using the
support information from the mapping of long reads. 

While the alignments of long reads on the complete genome can be easily
identified using BLASR~\cite{chaisson2012mapping}, alignments to shorter
contigs can be spurious. This is because we have to allow very short alignments
and cannot conclusively say if these are true alignments or accidental
alignments due to short repeats and similar-looking regions. Such alignments
generate ambiguous adjacencies between contigs and stop us from making high
confidence calls while scaffolding. Furthermore, as we see in
Figure~\ref{fig:repeatdist}, the longer assembled contigs tend to be unique in
the genome whereas the shorter contigs tend to repeat. If such short contigs
occur adjacent to each other in the genome, then using long reads to determine
the exact layout of all the contigs in the presence of spurious alignments
becomes a difficult problem.

\textbf{Contribution:} In this work, we present \textit{Cerulean}, a completely
automated hybrid assembly approach to produce a high quality scaffolds using
Illumina paired-end reads and PacBio long reads. Cerulean does not use the
short reads directly; instead it works with an \textit{assembly graph}
structure generated from short read data using existing assemblers. Assembly
graphs are graphs where nodes correspond to contigs of assembled short reads
and edges represent putative adjacencies of the contigs, but not confined to
overlapping contigs. Such assembly graph are commonly built using
overlap-layout-consensus and more recently with the de Bruijn graph paradigm.
The input to Cerulean includes: (i) the \textit{assembly graph} generated by
ABySS paired-end read assembler ~\cite{simpson2009abyss} constructed from
short reads (and it can be applied to graphs generated by overlap graph or de
Bruijn graph based assemblers as long as the graph has the desired format) and
(ii) the mapping of long reads to the assembled contigs. The output of Cerulean
is an simplified representation of the unentangled assembly graph. The
non-branching paths in this simplified graph correspond to the scaffolds in the
genome.

We recognize that multiple spurious alignments of long reads to short contigs
makes it a difficult problem to unentangle the assembly graph, especially when
the short contigs form densely connected substructures. The Cerulean algorithm
addresses this problem through an iterative framework to identify and extend
high confidence genomic paths. Cerulean initially operates with a simplified
representation of the assembly graph, which we call the \textit{skeleton graph}
(Figure~\ref{fig:skeletongraph}), consisting only of long contigs. We then
gradually improve the assembly by adding smaller contigs to the skeleton graph
in each iteration.

Our software produced higher quality assembly than the state-of-the-art
software for hybrid assembly (PacbioToCA, AHA) in much shorter running time and
lower memory usage. While assembly of Cerulean was significantly better than AHA
scaffolder; PacbioToCA required 8 hours to run on a supercomputer with 24
threads for a bacterial genome dataset with a total memory usage of 55GB and
temporary files of 300GB. In contrast, Cerulean finished within few minutes on a
single thread on a regular desktop with memory usage of 100MB and preprocessing
(ABySS, BLASR) taking less than an hour on a desktop computer. Starting with
N75 of 60 kbp of contigs generated by ABySS, scaffolds generates scaffolds with
N75 of 503 kbp as compared to 247 kbp for PacbioToCA and 106 kbp for AHA
scaffolder.

\section{Methods}

\textbf{Inputs:}
The inputs to Cerulean include (i) the assembly graph and contig sequences from
short read assembly (using ABySS or other assemblers) and (ii) alignments of
long reads to contigs from the assembly graph (using BLASR). The assembly graph
consists of one vertex for each contig and a conjugate vertex for its reverse
complement. The length of a vertex corresponds to the length of the contig. A
directed edge between two vertices indicates a putative adjacency between the
two vertices. For every directed edge, the conjugate edge is a directed edge
from the conjugate of its sink to the conjugate of its source. For each edge, we
define the length to be the offset between the end of source contig and start of
the sink contig. Thus, if the two contigs overlap, then the length is a negative
number; in case of a gap this length is a positive number. The size of the
overlap or the gap may depend on the short read assembler, e.g., if the
assembler just produces the de Bruijn graph, then the overlap is directly
determined by $k$-mer size, but many short read assemblers also implement
preliminary analysis of the graph structure and so the assembled contigs may
even overlap by a few thousand base pairs even though the paired-end insert
size is only few hundred based pairs. Henceforth, \textit{contigs} contigs
refer to DNA sequences assembled by the short read assembler and scaffolds
refer chain of contigs glued together using alignments of long reads to contigs.

\begin{figure}[tb]
\includegraphics[width=\textwidth]{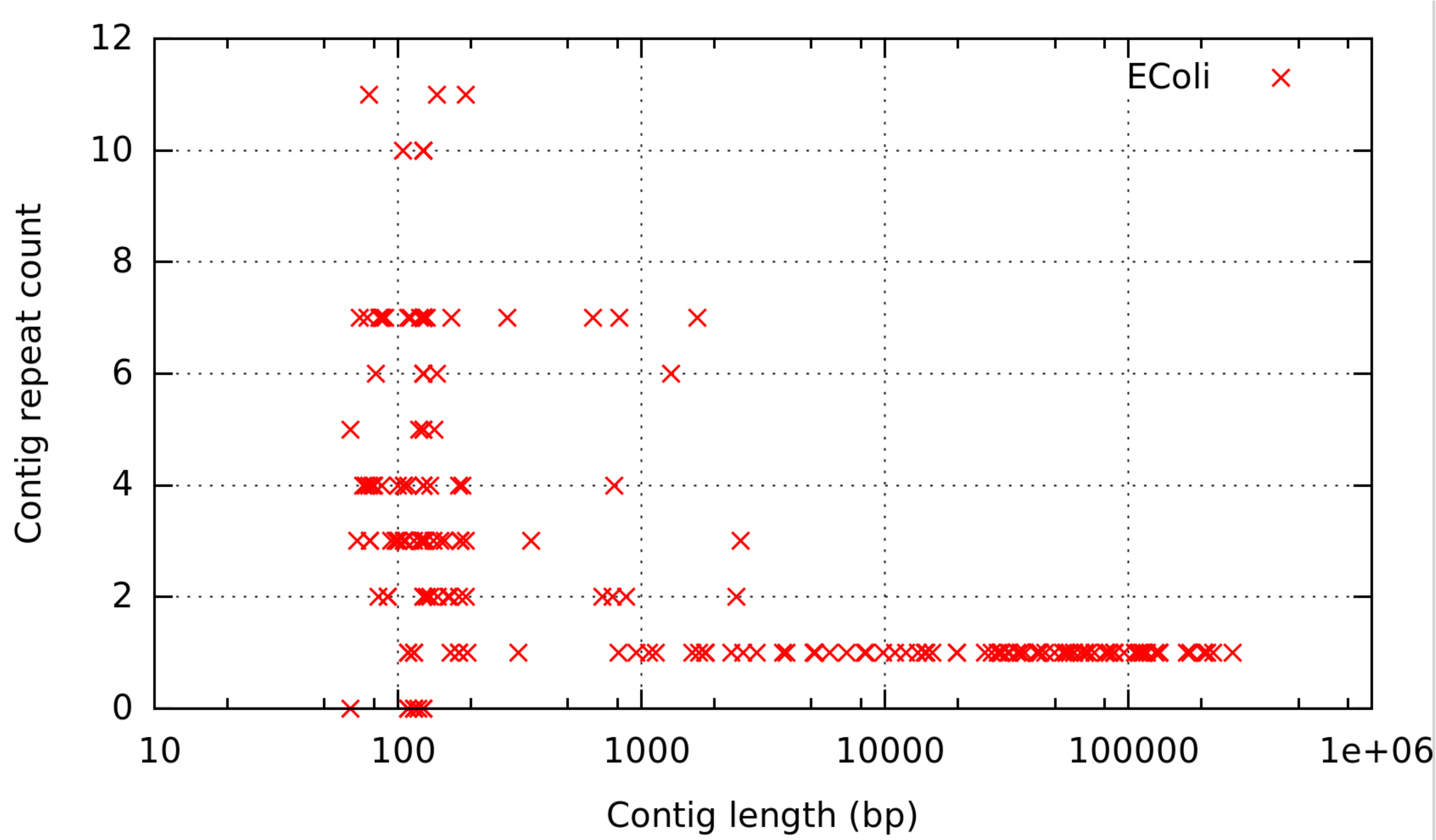}
\caption{Distribution of repeat count of contigs assembled from short reads}
\label{fig:repeatdist}
\end{figure}

\textbf{Pipeline:}
The contigs generated by the short paired-end read assembly have a large
distribution of lengths and some of these contigs repeat multiple times in the
reference. Most repeats tend to be short and there are very few long contigs
which occur multiple times in the reference as shown in
Figure~\ref{fig:repeatdist} for E. Coli dataset. Resolution of the assembly
graph in the presence of these short and repetitive contigs is difficult since
they create noise in mapping (spurious alignments) and may form dense
structures in the graph which is a major obstacle for the repeat resolution
procedure.

\begin{figure}
 \begin{center}
  \subfigure [(i) Assembly graph (ii) Skeleton graph retains only big vertices
(long contigs)]{
   \includegraphics[width=\textwidth]{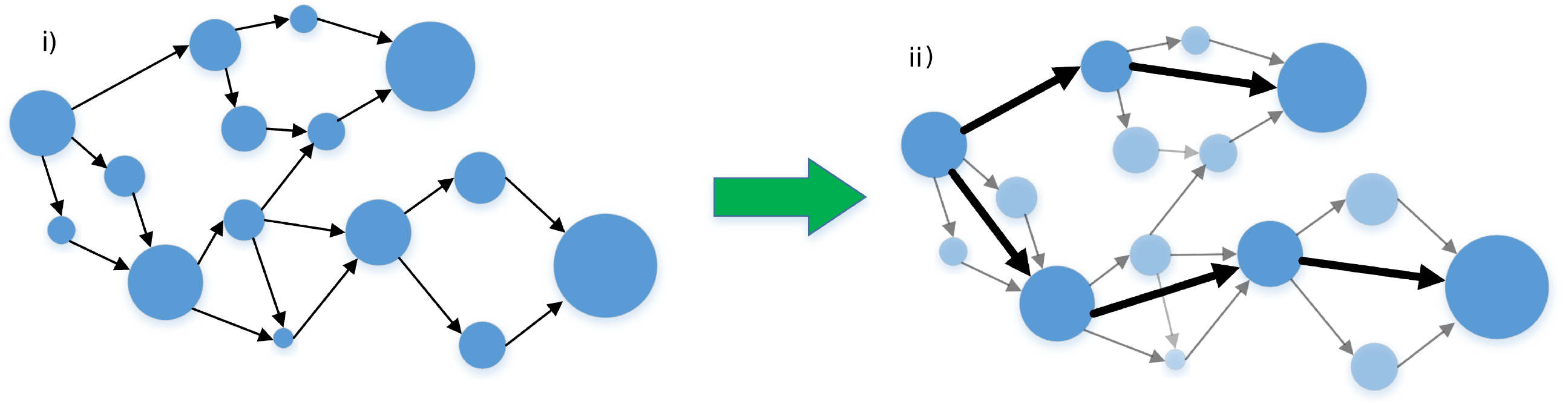}
   \label{fig:skeletongraph}
  }

 \subfigure [Iterative Pipeline for Cerulean: (i) Skeleton graph reconstruction
(Circle size proportional to contig length) (ii) Repeat resolution (iii) Gap
filling.]{
  \includegraphics[width=\linewidth]{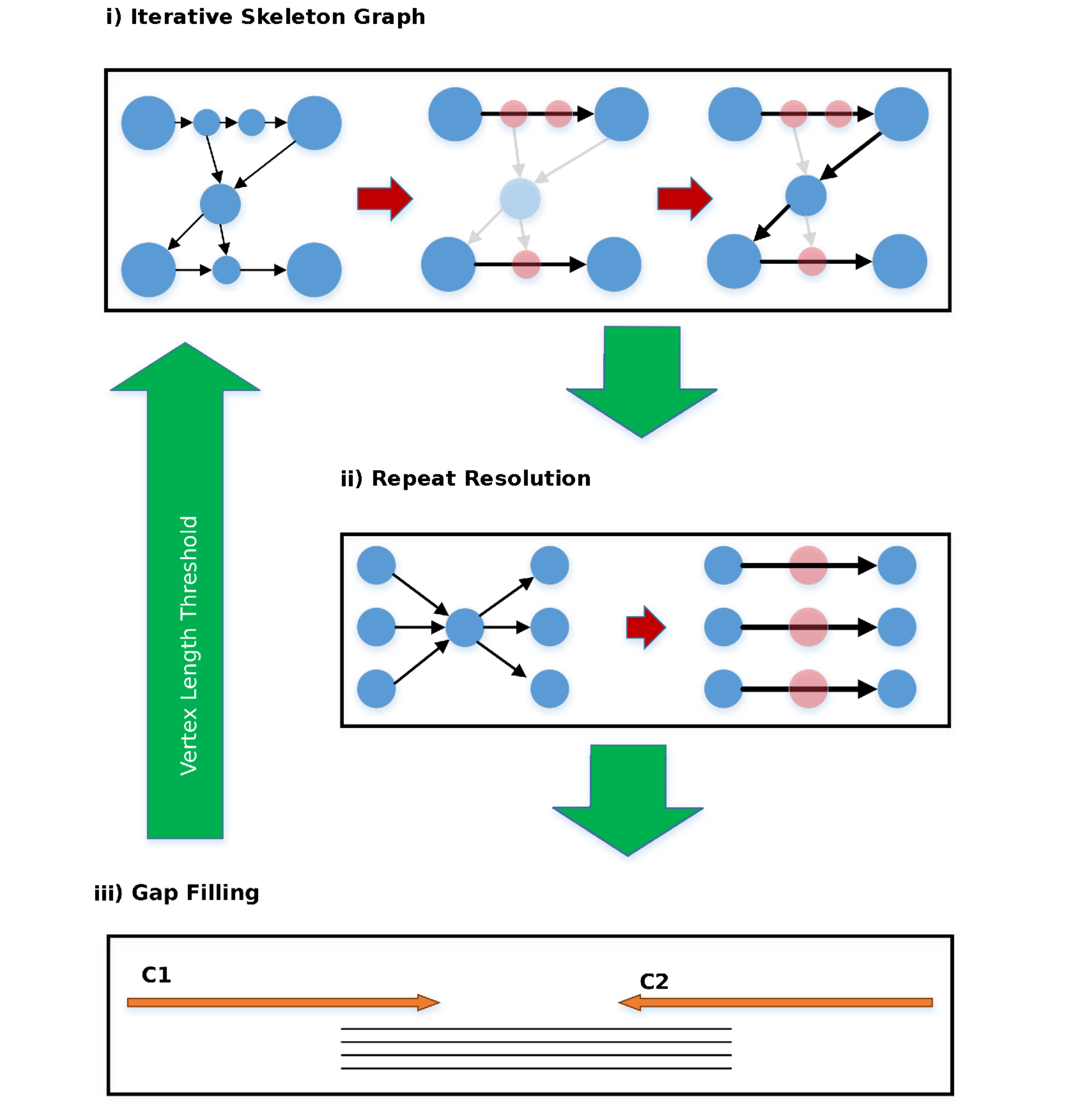}
  \label{fig:pipeline}

 }
 \caption{Skeleton graph representation and iterative construction of
approximate skeleton graph}
 \end{center}
\end{figure}

Our algorithm relies on the construction of a \emph{skeleton graph}
(Figure~\ref{fig:skeletongraph}) which is a simplified representation of the
assembly graph containing only long contigs. The edges in the skeleton graph
represent the putative genomic connections of these contigs. \textit{As we
shall see below, we include an edge in the skeleton graph only if there is
sufficient number of long reads that indicate the corresponding adjacency}.
Since the skeleton graph has a simple structure consisting only of long contigs,
mapping long reads to the skeleton graph has less noise and repeat resolution is
simpler. Our approach gradually improves the assembly by adding smaller contigs
in every iteration.

In each iteration, the algorithm goes through three
components (Figure~\ref{fig:pipeline}):
I) Skeleton graph
construction/extension; II) Repeat Resolution ; III) Gap bridging.

\textbf{Skeleton graph construction:} Given the assembly graph $G(V,E)$, genome $S$, a length threshold $L$, the
skeleton graph $SG(G, S, L) = G(V', E')$ has vertex set $V'$ containing only
vertices corresponding to contigs longer than $L$ in $V$. An edge $(v'_i,v'_j)
\in E'$ depicts a putative adjacent layout between the corresponding contigs in
the genome. The skeleton graph represents a simplification of the original
assembly graph by ignoring all intermediate short contigs in the assembly graph.
The short contigs that occur between consecutive long contigs in the genome are
implicitly included by annotating the relevant edges.

Since the skeleton graph is a simpler graph with long vertices, unambiguously
mapping the long reads to long contigs and resolving repeats in the skeleton
graph is more favorable than in the assembly graph. However, since the genome is
unknown, the skeleton graph can not always be constructed in its entirity.
Below, we construct an approximate version of skeleton graph using the
information from all long read alignments.

The first iteration of the skeleton graph is constructed by using only long
contigs from the assembly graph as vertices. Alignments of pairs of long contigs
to a long read imply a certain distance (overlap or gap) between the contigs
if these were true alignments. A directed edge is added between 2 vertices if
there exists a path (or edge) in the assembly graph that certifies the implied
distance within certain tolerance. The length of the edge is defined as
the distance inferred by following the path (in the assembly graph) rather than
the distance inferred from erroneous long reads. In case the adjacent contigs
overlap in the assembly graph, then the long reads need to span the entire
overlap to include a putative edge between the two contigs in the skeleton
graph.

We further refine this approximate graph by the following steps:

\emph{Read Count Threshold:} We should only keep those edges which have a
significant long read support. The length of the long reads is variable. So if
we want to resolve a long repeat or fill a large gap, then we will expect a
small number of long reads to span the entire repeat or gap. Thus, the expected
number of long reads that connect two contigs will depend on the coverage as
well as the distance between the two contigs and the read length distribution of
the long reads. We thus evaluate the significance of the number of long reads
supporting an edge in comparison to support for other competing edges. Our
criteria for adding new edges consists of three parts: (i) if the number of long
reads supporting an edge is greater than a high confidence threshold that edge
is certainly retained; (ii) if the number of supporting long reads is less than
a certain low confidence threshold, then such an edge is discarded; (iii) if
long read count is between these thresholds, an edge is included either if it
is the only outgoing/incoming edge for the source/sink, or if the read count for
the edge is significantly higher than other edges incident on source or sink.

\begin{figure}
\begin{center}

 \includegraphics[width=0.5\linewidth]{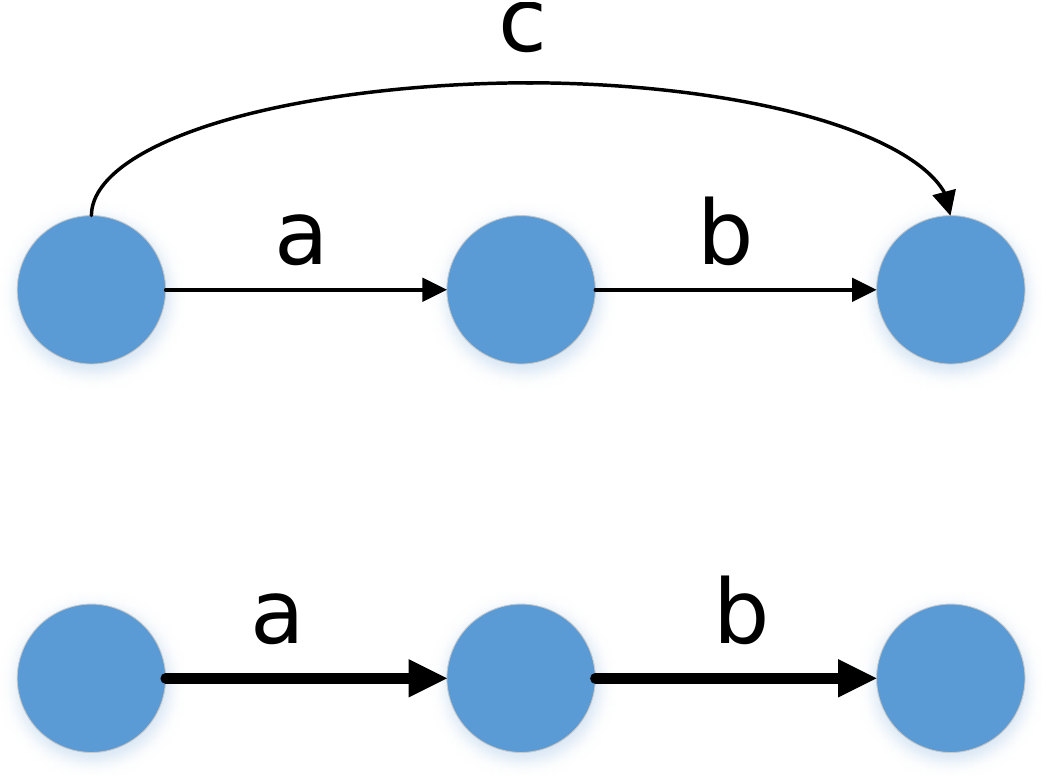}
 \caption{Length-sensitive transitive edge reduction along a non-branching
path.}

 \end{center}
 \label{fig:transitivereduction}
\end{figure}

\emph{Length-sensitive Transitive Edge Reduction:} We can identify the high
confidence scaffolds by looking at non-branching paths in the skeleton graph.
However, some chains of contigs which ideally should form non-branching paths in
the graph may get connected to vertices beyond their immediate neighbours. Such
cases can happen when some long reads do not align to the intermediate vertex
due to errors. For identification of non-branching paths, we remove the
transitive edges which create the false impression of a branch
(Figure~\ref{fig:transitivereduction}). An edge is defined to be transitive if
there is an alternative path from source to sink implying the same distance
offset.

\textbf{Repeat Resolution:} The repeat resolution procedure is illustrated in
Figure~\ref{fig:pipeline}(ii). When we remove the transitive edges to identify
simple paths, we may lose valuable information from long reads that span repeats
appearing as branching nodes in the graph. A branching node will have multiple
incoming edges on one side on multiple outgoing edges on the other. The
corresponding incoming and outgoing vertices form a bipartite structure as seen
in Figure~\ref{fig:pipeline}(ii). We resolve such repeats by matching pairs of
vertices (not necessarily all pairs) in this bipartite structure by looking at
reads spanning the repeat. Matched pairs which can be joined unambiguously are
then connected by a direct edge (with annotation) and the edges to the
intermediate vertex are removed. We do repeat resolution in its simplest form of
spanning only one contig at a time. This works if the repeating contig is a
maximal repeat. It is the common case for most contigs in our procedure to be
maximal repeats since we deal with only long contigs initially. But as we scale
to larger genomes with more complex repeat structures of shorter contigs, we
need to implement more powerful strategies for spanning repeats.

\textbf{Gap Bridging:} The gap bridging procedure is illustrated in
Figure~\ref{fig:pipeline}(iii). After using all information from connections in
the graph to identify the scaffolds, we observe that certain paths terminate as
there are no further edges to extend these paths. At this point, we relax the
constraint that edges in the skeleton graph should correspond to existing paths
of the assembly graph. We can identify possible ways to extend a scaffold
by looking at long reads that align from the end of that scaffold to the end of
another scaffold. If this is the only possible way to extend either of these
scaffolds, then we use these long read alignments to unambiguously bridge the
gap between these paths (scaffolds) by adding a new edge between their terminal
vertices.

\textbf{Iteration on Vertex Length:} After we have inferred all possible
scaffolds from the long contig skeleton, we have the list of all simple paths 
from this skeleton. Now we can decrease the vertex length threshold according to
a vertex length schedule. We add the new short contig vertices (longer than the
new lower threshold) either if they are connected to the end vertices of the
simple paths from the previous vertex length iteration, or if they are connected
to other short vertices. Thus non-terminal non-branching long contigs in
scaffolds from one iteration are untouched in the next iteration. Then we
iteratively go through the above steps of transitive edge reduction, repeat
resolution and gap bridging.

\textbf{Final assembly:} After the completion of all iterations through the
vertex length schedule we have our final approximation of the skeleton graph.
The simple paths represent our final scaffolds. The edges of these simple paths
can either be directed edges in the original assembly graph, or they can be
annotated with either path traversals in the assembly graph by the set of
long reads bridging a gap. Thus, the final scaffolds can be inferred from these
annotated paths on the corresponding sequence of reads that are used for gap
filling. Those vertices which are not included in the scaffolds can be inferred
as independent contigs.

\section{Results}

We tested our software for the Escherichia Coli bacterial genome (strain:
K12, isolate: MG1665). The short~\cite{illumina2013ecoli} and long~\cite{pacbio2013ecoli} read datasets
were obtained from the samples provided by Illumina and Pacific Biosciences
respectively as described in
Table~\ref{tab:dataset}.

\begin{table}[b]
 \caption{Details of sequencing data used for assembly}
\begin{tabular}{| l || c | c |}
 \hline
 \textbf{Platform} 	& \textbf{Illumina Hiseq}	& \textbf{PacBio RS} \\
\hline \hline
 \textbf{Coverage} 	& 400X			& 30x       \\ \hline
 \textbf{Read Length}	& 151bpX2 (insert size 300bp) & N50: 5900,
Largest:19416\\ \hline
 \textbf{Number of reads} & 11 million                   & 75152\\ \hline
\end{tabular}
 \label{tab:dataset}
\end{table}

\textit{Short read assembly:} We assembled the short read contigs using the
ABySS paired-end assembler with $k$-mer size of 64 base pairs. The
computational resources and assembly results are mentioned in
Table~\ref{tab:compres} and Table~\ref{tab:results} respectively.

\textit{Mapping long reads to contigs:} We mapped the long reads to the ABySS
assembled contigs using BLASR with minimum percentage identity of 70\%. 

\begin{figure}
 \includegraphics[width=\linewidth]{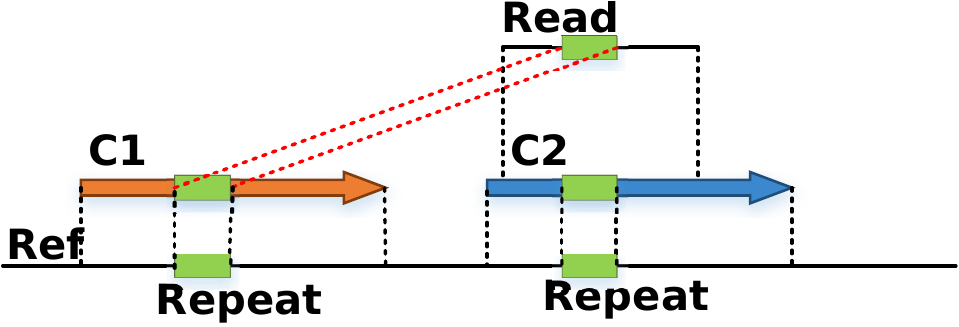}
 \caption{Spurious alignments of long reads due to short repeats are filtered.}
 \label{fig:spuriousmapping}
\end{figure}

\textit{Filtering spurious alignments:} There are many short alignments from
long reads to contigs due to short repeats contained within contigs as shown in
Figure \ref{fig:spuriousmapping}. Reads mapping to multiple contigs do not
necessarily imply adjacency and need to be filtered. We classify an alignment as
long alignment if the unaligned overhang (i.e. length of unaligned portion of
the read which ideally should have mapped in case of a true alignment of
unerroneous read) is less than 30\% of the ideal alignment length (i.e. sum of
unaligned overhang and aligned portion). Henceforth, when we refer to alignment
of a long read it means it satisfies the criteria for long alignment.

\textit{Cerulean scaffolding:} We used the ABySS assembly graph and filtered
BLASR alignments to generate the scaffolds. The vertex length schedule we used
is 2048 bp, 1024 bp, 512 bp, 256 bp and 0 bp.

\textit{PacbioToCA:} We compare our results to the assembly generated
by the alternative approach of assembling error-corrected long reads using
PacbioToCA.

\textit{AHA scaffolder:} We also tested the results of the AHA
scaffolder using the ABySS assembled contigs and PacBio long reads as inputs.

\textit{ALLPATHS-LG:} We did not test ALLPATHS-LG since it requires jumping
libraries.

\begin{figure}
 \begin{center}
  \subfigure[Cumulative length distribution of all contigs/scaffolds
arranged in decreasing order of length]{
  \includegraphics[width=\linewidth]{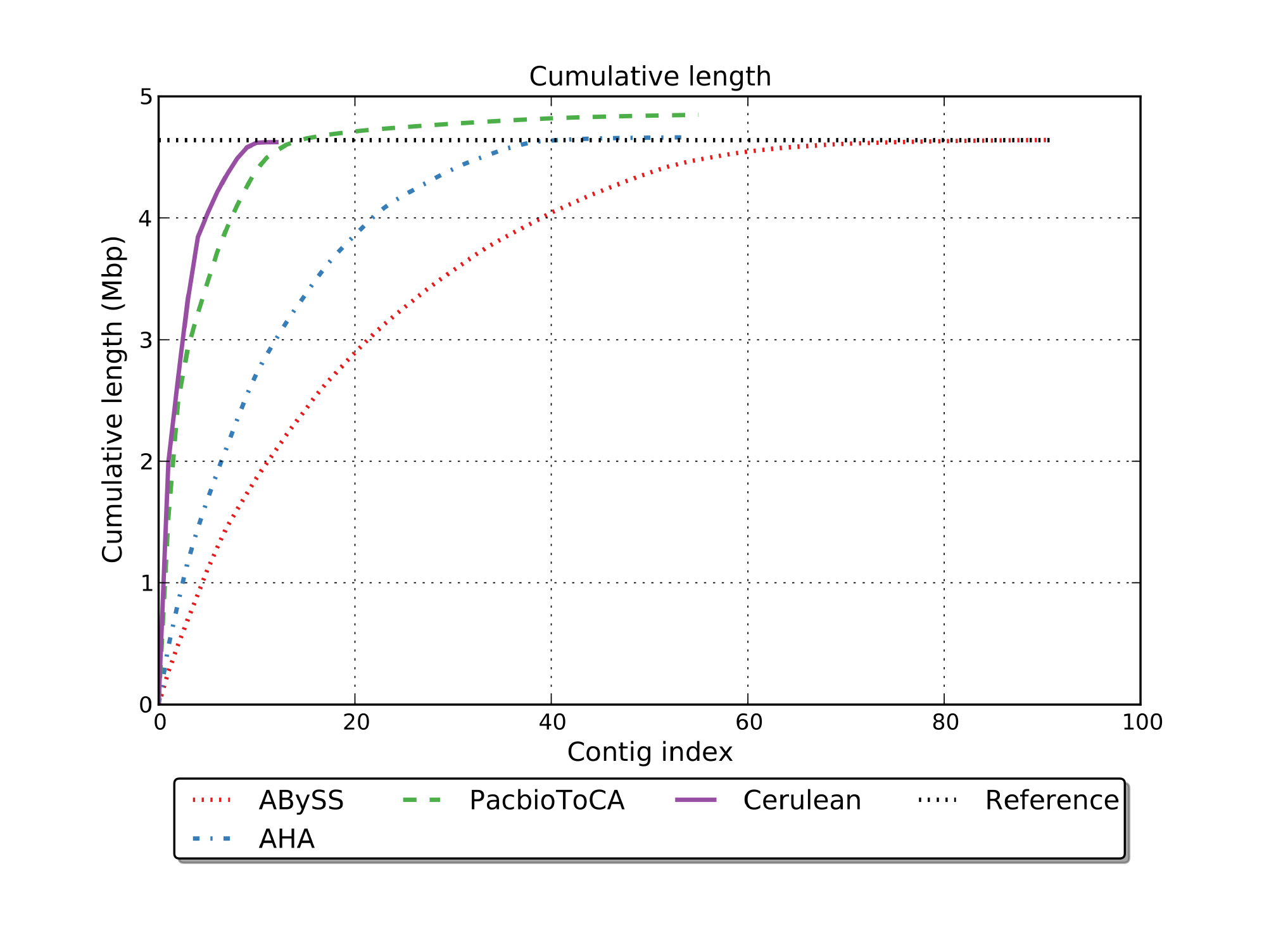}
  \label{fig:cumulative}
 }

 \subfigure [Nx length of all contigs/scaffolds of scaffolds arranged in
decreasing order of length]{
  \includegraphics[width=\linewidth]{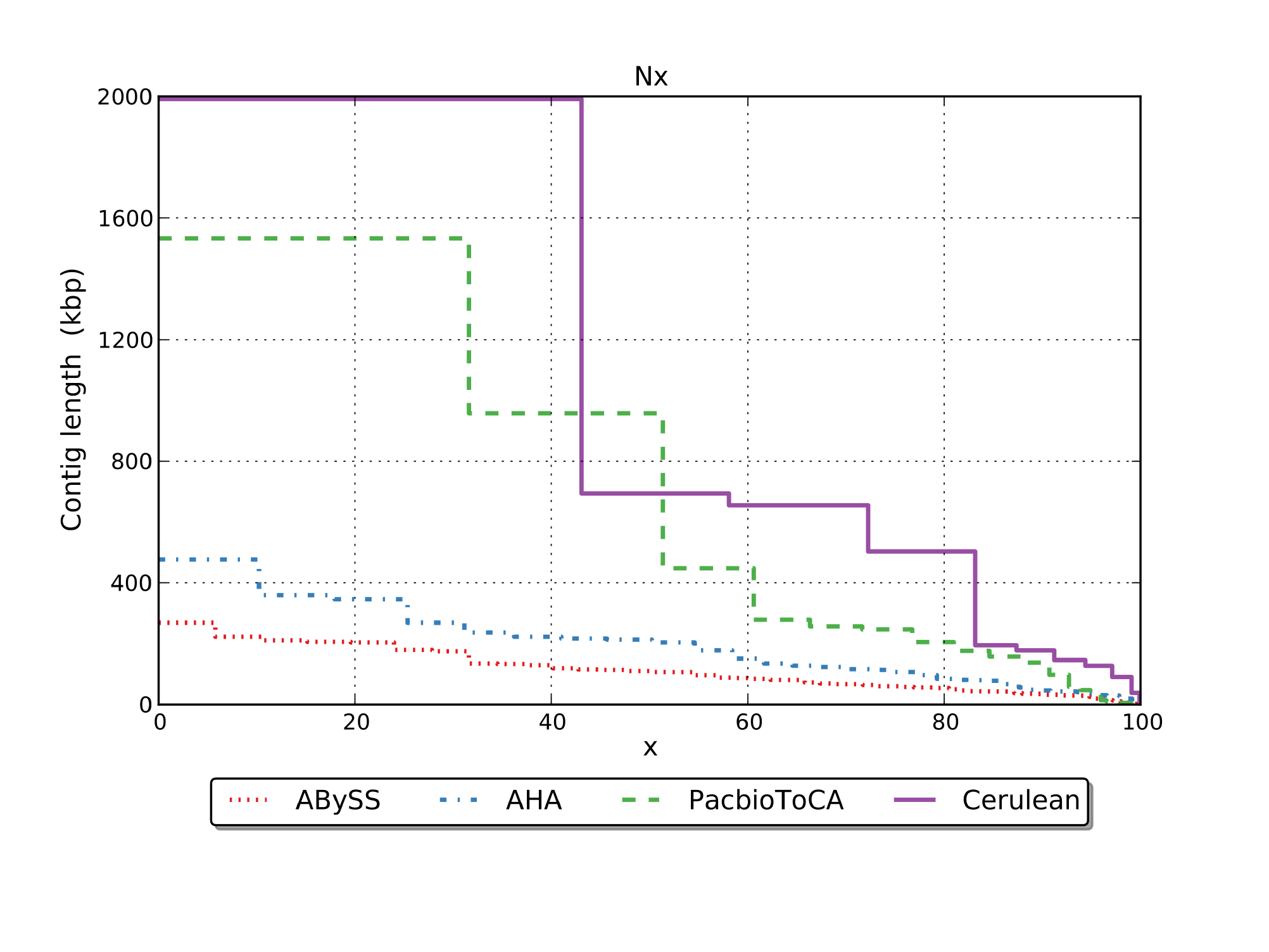}
  \label{fig:Nx}
 }
 \caption{Comparison of lengths of contigs/scaffolds generated by various
approaches}
 \end{center}
\end{figure}

 The length distribution comparison of the assembled contigs is displayed in
Figures~\ref{fig:cumulative} and~\ref{fig:Nx}. We can see that the N50 values
for the PacbioToCA assembly is determined solely by the first 2 contigs, but the
contig length drops drastically after that giving a very low N75 value of 273
Kbp as compared to N50 of 957 Kbp. The length of the scaffolds generated by
Cerulean falls much slower giving a significantly better N75 of 503 Kbp
comparable to the N50 length 694 Kbp. Figure~\ref{fig:cumulative} also shows
that the total assembled length for PacbioToCA is significantly larger than the
genome length.

\begin{table}
  \caption{Computational resources for hybrid assembly}
  \label{tab:compres}
\begin{tabular}{| l || c | c | c || c | c |}
 \hline
 \textbf{Software} 	& \textbf{ABySS}	& \textbf{BLASR}  &
\textbf{Cerulean} & \textbf{PacbioToCA} & \textbf{AHA}\\
\hline \hline
 \textbf{Number of threads} 	& 1 & 8 & 1 & 24 & 4      \\ \hline
 \textbf{Peak memory usage}	& $<$4 GB & 300 MB & 100 MB & 55 GB &
300 MB\\
\hline
 \textbf{Runtime} & $<$30 mins & $<$30 mins & 2 mins & 12
hours & 2 hours\\
\hline
 \textbf{Temporary files} & 75 MB & 8 MB & 5 MB & 300 GB & 500 MB\\
\hline
\end{tabular}
\end{table}

\begin{table}
  \caption{Assembly statistics for hybrid assembly analyzed using
QuAsT~\cite{schmutz2004quality}. We have separately validated the sequential
order and offsets of all long contigs forming the scaffold by mapping them to
the reference. We also confirmed that all 4 reported misassemblies for ABySS +
Cerulean were actually local misassemblies of a small length ($< 1Kbp$) and 1
fake misassembly due to circular genome. However, a significant number of
misassemblies in PacbioToCA and AHA involved relocations/inversions of
significant number of contigs longer than 1Kbp and as large as 30 Kbp and 19
Kbp respectively.}
  \label{tab:results}
\begin{tabular}{| l || c || c | c || c | c |}
 \hline
 \textbf{Software} & \textbf{Reference}	& \textbf{ABySS}	&
\textbf{Cerulean} & \textbf{PacbioToCA }  & AHA\\
\hline \hline
 \textbf{\# contigs} & 1	& 199 & 21 & 55 & 54 \\ \hline
\textbf{\# contigs $>$ 1000bp} & 1	& 83 & 11 & 55 & 48 \\ \hline
 \textbf{N50} & 4639675	& 110Kbp &  694KBp & 950Kbp & 213 Kbp\\ \hline
 \textbf{N75} & 4639675 & 64KBp & 507KBp & 247 KBp & 107 Kbp\\ \hline
\textbf {Largest contig length} & 4639675 & 268969 & 1991897 &1533073 & 477080\\
\hline
\textbf {Total length} & 4639675 & 4849724 & 4625935 & 4641287 & 4663300\\
\hline
 \textbf{\#misassemblies} & - & 3 & 4 & 22 & 11\\
\hline
\end{tabular}
\end{table}

Analysis of the final set of contigs indicated that all the 11 long contigs were
essentially separated by just 2 non-exact repeats in the reference of lengths
(2500 bp and 4300 bp). If we are more aggressive in resolving these repeats,
then this approach has the potential to retrieve the entire genome as a
single contig. Our conservative adjacency calls did not resolve these repeats in
order to retain the accuracy of the assembly. We chose not to make more
aggressive decisions in repeat resolution in the dataset because in the case of
just 2 repeats, it is easy to implement a scheme that will overfit the data and
not scale to other genomes when running in a fully automated setting.

\section{Discussions:}

Cerulean has a very low resource usage and high accuracy of assembled scaffolds.
This makes a very strong case for scaling this approach to larger genomes. The
algorithm in its current state focuses on making decisions based on very simple
building blocks one at a time. This makes it possible for us to make low risk
decisions towards a high accuracy assembly for simple bacterial genomes.
However, when analyzing datasets from larger complex genome, we have no
prior knowledge of the structure of the repeats and the layout of the
contigs generated by short read assemblers. So there are cases where the
scaffolding algorithm may not be able to distinguish between a true adjacency
signal and a false adjacency signal. In most cases, this will simply stop the
algorithm from extending a scaffold due to branching. However,
we cannot conclusively rule out the possibility of producing
other side effects for every decision made by the algorithm. We also need to
acknowledge the fact that we are currently dealing with small bacterial genomes
for which we can easily obtain high and more or less uniform coverage for short
reads. So far we rely completely on the short read assembler to generate the
initial contigs. However, for larger genomes, variable coverage caused due to
sequencing bias combined with decisions made by the short read assembler can
cause misassembled contigs to start with. In this case, the scaffolder will
benefit from not assuming the assembled contigs as ground truth, but actually
testing for misassemblies by the short read assembler. 

However, the framework of gradual inclusion of complexity does provide us
with the opportunity to tackle even more complex genomes in a systematic
fashion. Here we discuss a few of these cases where this framework is useful.
When extending the scaffolds consisting of large contigs, we aim to extend them
unambiguously with the inherent assumption that the larger contigs are usually
unique in the reference. We can increase the confidence in this assumption by
considering the coverage information of contigs to estimate the repeat counts of
contigs and make our decision for extension based on this
assumption. Furthermore, if in one iteration of the vertex length schedule, we
find an unambiguous way of extending a scaffold, it does not necessarily imply
that it is the only way  to extend and there can possibly be other ways to
extend which require looking at shorter vertices for bridging. One way to
address problem is by using the vertex length threshold as a soft cutoff
rather than a hard cutoff to allow shorter contigs to compete with the larger
contigs if they have very good support from the reads and graph structure.

\begin{figure}[bt]
  \includegraphics[width=\textwidth]{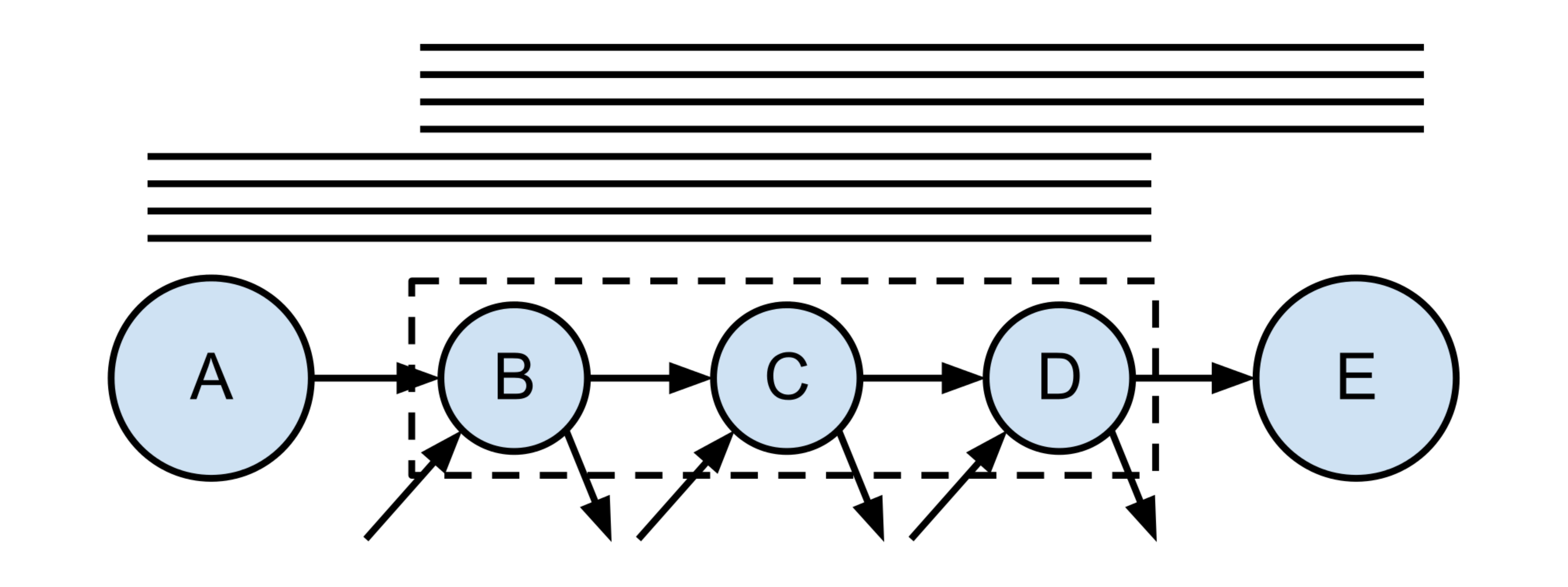}
 \caption{Example for resolving multiple consecutive repeats. Contigs A and E
are not connected directly by long reads. The intermediate vertices B, C
and D all have branches however, the triplet B $\rightarrow$ C $\rightarrow$ D
only occurs at a unique location. Thus reads which span all 3 vertices B, C and
D simultaneously can be treated as aligning to one large combined contig
(3-gram) and used to extend the scaffold from A to E.}
  \label{fig:multrepextension}
\end{figure}

In Cerulean, we  bypass the problems of complex repetitive structures
in graph, by only looking at uniquely occurring long vertices. We resolve only
one branching vertex at a time under the inherent assumption that most
long repeats are maximal repeats isolated from other repeats of comparable
length. Larger genomes certainly violate this assumption to a large extent. Thus
we may have to connect scaffolds that are separated by multiple branching
vertices of comparable lengths. There is also the opportunity to exploit the
structure of the graph in extending the scaffolds as shown in Figure
\ref{fig:multrepextension}. A path tracing approach can help us extend scaffolds
across such multiple branching vertices. Path tracing is a non-trivial problem
in big graphs, but we can use path tracing in the context of our incremental
framework to solve specific small problems.

Finally there are techniques we can use to improve our ability to extend
scaffolds or add more edges to the skeleton graph. An edge in the skeleton graph
can correspond to a walk in the assembly graph. The input mapping from long
reads to contigs may miss some alignments from the long reads to intermediate
contigs due to high error rate. In such a case, we can perform a more informed
search for such a walk by looking at local alignments of reads along neighoring
contigs. If we can identify true chains of small contigs, then we can
rerun the mapping the reads to the
concatenated sequences of these chains and use these mappings while extending
existing scaffolds. PBJelly~\cite{english2012mind} has displayed that gaps in
scaffolds can be filled with high accuracy using the mapping reads. Thus while
retrieving the the intermediate sequences of scaffolds, we can use a combination
of long reads and assembled contigs to bridge the long contigs. After
we have finished making the most conservative calls using the assembled short
read contigs and filled in all gaps using the pairwise alignments of the long
reads, we can be more ambitious and bridge the gaps between the scaffolds
with targeted assembly of long reads.

In conclusion, we present a hybrid assembly approach that is both
computationally effective and produces high quality assemblies. Our algorithm
first operates with a simplified version of the assembly graph consisting only
of long contigs and gradually improve the assembly by adding smaller contigs in
each iteration. In contrast to the state-of-the-art long reads error 
correction technique, which requires high computational resources and long
running time on a supercomputer even for bacterial genome datasets, our software
can produce comparable assembly using only a standard desktop in a short running
time.

\section*{Acknowledgments}
The authors will like to sincerely thank Pavel Pevzner and Glenn Tesler for
their insightful comments. V.D. and V.B. were supported in part by NIH grants
5RO1-HG004962 and U54 HL108460, and by the NSF grant NSF-CCF-1115206. S.P. was
supported in part by NIH grant 3P41RR024851-02S1.

\section*{Disclosure Statement}
No competing financial interests exist.

\bibliography{Cerulean}

\begin{thebibliography}{10}

\bibitem{koren2012hybrid}
Koren, S., Schatz, M.C., Walenz, B.P., Martin, J., Howard, J.T., Ganapathy, G.,
  Wang, Z., Rasko, D.A., McCombie, W.R., Jarvis, E.D.,  et~al.:
\newblock Hybrid error correction and de novo assembly of single-molecule
  sequencing reads.
\newblock Nature biotechnology \textbf{30}(7) (2012)  693--700

\bibitem{staden1979strategy}
Staden, R.:
\newblock A strategy of dna sequencing employing computer programs.
\newblock Nucleic acids research \textbf{6}(7) (1979)  2601--2610

\bibitem{myers2005fragment}
Myers, E.W.:
\newblock The fragment assembly string graph.
\newblock Bioinformatics \textbf{21}(suppl 2) (2005)  ii79--ii85

\bibitem{myers2000whole}
Myers, E.W., Sutton, G.G., Delcher, A.L., Dew, I.M., Fasulo, D.P., Flanigan,
  M.J., Kravitz, S.A., Mobarry, C.M., Reinert, K.H., Remington, K.A.,  et~al.:
\newblock A whole-genome assembly of drosophila.
\newblock Science \textbf{287}(5461) (2000)  2196--2204

\bibitem{simpson2012efficient}
Simpson, J.T., Durbin, R.:
\newblock Efficient de novo assembly of large genomes using compressed data
  structures.
\newblock Genome Research \textbf{22}(3) (2012)  549--556

\bibitem{idury1995new}
Idury, R.M., Waterman, M.S.:
\newblock A new algorithm for dna sequence assembly.
\newblock Journal of Computational Biology \textbf{2}(2) (1995)  291--306

\bibitem{pevzner2001eulerian}
Pevzner, P.A., Tang, H., Waterman, M.S.:
\newblock An eulerian path approach to dna fragment assembly.
\newblock Proceedings of the National Academy of Sciences \textbf{98}(17)
  (2001)  9748--9753

\bibitem{chaisson2008short}
Chaisson, M.J., Pevzner, P.A.:
\newblock Short read fragment assembly of bacterial genomes.
\newblock Genome research \textbf{18}(2) (2008)  324--330

\bibitem{simpson2009abyss}
Simpson, J.T., Wong, K., Jackman, S.D., Schein, J.E., Jones, S.J., Birol,
  {\.I}.:
\newblock Abyss: a parallel assembler for short read sequence data.
\newblock Genome research \textbf{19}(6) (2009)  1117--1123

\bibitem{zerbino2008velvet}
Zerbino, D.R., Birney, E.:
\newblock Velvet: algorithms for de novo short read assembly using de bruijn
  graphs.
\newblock Genome research \textbf{18}(5) (2008)  821--829

\bibitem{eisenstein2013companies}
Eisenstein, M.:
\newblock Companies' going long'generate sequencing buzz at marco island.
\newblock Nature biotechnology \textbf{31}(4) (2013)  265--266

\bibitem{waldbieser2013production}
Waldbieser, G.:
\newblock Production of long (1.5 kb--15.0 kb), accurate, dna sequencing reads
  using an illumina hiseq2000 to support de novo assembly of the blue catfish
  genome.
\newblock In: Plant and Animal Genome XXI Conference, Plant and Animal Genome
  (2013)

\bibitem{chin2013nonhybrid}
Chin, C.S., Alexander, D.H., Marks, P., Klammer, A.A., Drake, J., Heiner, C.,
  Clum, A., Copeland, A., Huddleston, J., Eichler, E.E.,  et~al.:
\newblock Nonhybrid, finished microbial genome assemblies from long-read smrt
  sequencing data.
\newblock Nature methods (2013)

\bibitem{au2012improving}
Au, K.F., Underwood, J.G., Lee, L., Wong, W.H.:
\newblock Improving pacbio long read accuracy by short read alignment.
\newblock PLoS One \textbf{7}(10) (2012)  e46679

\bibitem{hercus2009novocraft}
Hercus, C.:
\newblock Novocraft short read alignment package.
\newblock Website http://www. novocraft. com (2009)

\bibitem{wu2005gmap}
Wu, T.D., Watanabe, C.K.:
\newblock Gmap: a genomic mapping and alignment program for mrna and est
  sequences.
\newblock Bioinformatics \textbf{21}(9) (2005)  1859--1875

\bibitem{bashir2012hybrid}
Bashir, A., Klammer, A.A., Robins, W.P., Chin, C.S., Webster, D., Paxinos, E.,
  Hsu, D., Ashby, M., Wang, S., Peluso, P.,  et~al.:
\newblock A hybrid approach for the automated finishing of bacterial genomes.
\newblock Nature biotechnology (2012)

\bibitem{ribeiro2012finished}
Ribeiro, F.J., Przybylski, D., Yin, S., Sharpe, T., Gnerre, S., Abouelleil, A.,
  Berlin, A.M., Montmayeur, A., Shea, T.P., Walker, B.J.,  et~al.:
\newblock Finished bacterial genomes from shotgun sequence data.
\newblock Genome research \textbf{22}(11) (2012)  2270--2277

\bibitem{chaisson2012mapping}
Chaisson, M.J., Tesler, G.:
\newblock Mapping single molecule sequencing reads using basic local alignment
  with successive refinement (blasr): application and theory.
\newblock BMC bioinformatics \textbf{13}(1) (2012)  238

\bibitem{illumina2013ecoli}
:
\newblock {E.Coli MG1655 Illumina HiSeq2000 sequencing dataset}.
\newblock
  \url{ftp://webdata:webdata@ussd-ftp.illumina.com/Data/SequencingRuns/MG1655/MiSeq_Ecoli_MG1655_110721_PF.bam}
  (2013) [Online; accessed 24-June-2013].

\bibitem{pacbio2013ecoli}
:
\newblock {E.Coli K12 MG1655 Pacbio RS sequencing dataset}.
\newblock
  \url{http://files.pacb.com/datasets/primary-analysis/e-coli-k12/1.3.0/e-coli-k12-mg1655-raw-reads-1.3.0.tgz}
  (2013) [Online; accessed 24-June-2013].

\bibitem{schmutz2004quality}
Schmutz, J., Wheeler, J., Grimwood, J., Dickson, M., Yang, J., Caoile, C.,
  Bajorek, E., Black, S., Chan, Y.M., Denys, M.,  et~al.:
\newblock Quality assessment of the human genome sequence.
\newblock Nature \textbf{429}(6990) (2004)  365--368

\bibitem{english2012mind}
English, A.C., Richards, S., Han, Y., Wang, M., Vee, V., Qu, J., Qin, X.,
  Muzny, D.M., Reid, J.G., Worley, K.C.,  et~al.:
\newblock Mind the gap: Upgrading genomes with pacific biosciences rs long-read
  sequencing technology.
\newblock PloS one \textbf{7}(11) (2012)  e47768

\end{thebibliography}

\end{document}